\documentclass[aps,prl,twocolumn,preprintnumbers,superscriptaddress,groupedaddress,nofootinbib]{revtex4}  
\usepackage{bm}
\usepackage{amssymb}
\usepackage{amsmath,shuffle}

\allowdisplaybreaks

\usepackage{tikz}
\usetikzlibrary{calc} \usetikzlibrary{patterns} \usetikzlibrary{decorations.pathreplacing}
\usetikzlibrary{decorations.markings} \usetikzlibrary{decorations.pathmorphing} \usetikzlibrary{positioning}

\def\gg#1(#2,#3){g^{(#1)}_{#2#3}}

\def\eqref#1{(\ref{#1})}
\def\beq{\begin{equation}}
\def\eeq{\end{equation}}

\def\({\left(}
\def\){\right)}

\begin{document}

\title{Double-Copy Structure of One-Loop Open-String Amplitudes}

\author{Carlos R. Mafra}
\email{c.r.mafra@soton.ac.uk}
\affiliation{STAG Research Centre and Mathematical Sciences, University of Southampton, 
Highfield, Southampton SO17 1BJ, UK}
\author{Oliver Schlotterer}
\email{olivers@aei.mpg.de}
\affiliation{Max--Planck--Institut f\"ur Gravitationsphysik, Albert--Einstein--Institut, 14476 Potsdam, Germany}
\affiliation{Perimeter Institute for Theoretical Physics, Waterloo, ON N2L 2Y5, Canada}

\date{\today}

\begin{abstract}
In this Letter, we provide evidence for a new double-copy structure in
one-loop amplitudes of the open superstring. Their integrands
with respect to the moduli space of genus-one surfaces are cast into a form
where gauge-invariant kinematic factors and certain functions of
the punctures---so-called generalized elliptic integrands---enter on 
completely symmetric footing. In particular,
replacing the generalized elliptic integrands by a second copy of
kinematic factors maps one-loop open-string correlators to
gravitational matrix elements of the higher-curvature operator $R^4$.
\end{abstract}

\maketitle


\noindent
{\it \bf Introduction}. Recent investigations of scattering amplitudes revealed a variety of
hidden relations between field theories of seemingly unrelated
particle content. The oldest and possibly most prominent example of
such connections is the double-copy structure of gravity
\cite{Kawai:1985xq, Bern:2008qjA, Bern:2008qjB} whose scattering amplitudes can be 
reduced to squares of gauge-theory building blocks. This kind of
double copy is geometrically intuitive from the realization of
gravitons and gauge bosons as vibration modes of closed and open
strings, respectively. Its first explicit realization at the level
of scattering amplitudes in string theory was pinpointed by Kawai,
Lewellen and Tye (KLT) in 1985~\cite{Kawai:1985xq}.
  
The first loop-level generalization of the gravitational double copy
was found by Bern, Carrasco and Johansson (BCJ) \cite{Bern:2008qjB}:
Gauge-theory ingredients in a suitable gauge can be conjecturally
squared to gravitational loop integrands at the level of cubic
diagrams. The gauge dependence of the BCJ construction has been
recently bypassed through a generalized double copy
\cite{Bern:2017yxu} -- see \cite{Bern:2017ucb} for an impressive
five-loop application -- and a one-loop KLT
formula in field theory \cite{He:2016mzd}.

It has been recently discovered that tree-level amplitudes of the
{\it open} superstring admit a double-copy representation
\cite{Mafra:2011nv} which mimics the
field-theory version of the KLT formula \cite{Broedel:2013tta}: Gauge-theory trees are
double copied with moduli-space integrals whose expansion in the inverse string tension
$\alpha'$ suggests an interpretation as scattering amplitudes in effective scalar field 
theories \cite{Carrasco:2016ldy}.

One-loop open-string amplitudes exhibit two sorts of
invariances that are intertwined through a similar double-copy structure:
While gauge invariance is also required for field-theory amplitudes, 
string-theory correlators defined over a Riemann surface of genus one must be 
additionally invariant under monodromy variations, i.e.\ transporting 
their punctures around the homology cycles.

In this Letter, we introduce a one-to-one map between
gauge-invariant kinematic factors of the external states and
doubly periodic functions on genus-one Riemann surfaces, and the latter
will be traced back to so-called generalized elliptic integrands. The 
examples given up to six points provide evidence for a double-copy 
structure in one-loop open-string amplitudes. In particular, when the
generalized elliptic integrands are double copied to
their gauge-invariant kinematic counterparts, we obtain
gravitational tree-level matrix elements: Those with a single insertion 
of the higher-curvature operator $R^4$ from an effective Lagrangian
$\sim R+ R^4$ along with its supersymmetrization. 

The results of this 
Letter yield the first manifestly supersymmetric representations of seven-point 
integrands for open- and closed-string one-loop amplitudes, and we will 
report on cross-checks and higher-multiplicity results in~\cite{wip}.



\vspace{-0.2cm}

\vspace{-.6em}
\begin{figure}[h]
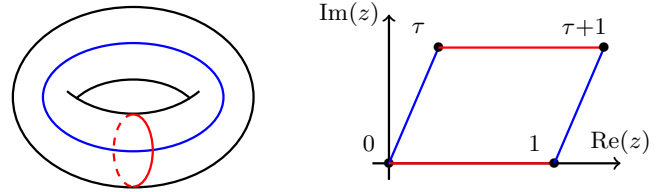

\begin{center}
\tikzpicture[scale=0.4,line width=0.30mm]
\draw(0,0) ellipse  (4cm and 3cm);
\draw(-2.2,0.2) .. controls (-1,-0.8) and (1,-0.8) .. (2.2,0.2);
\draw(-1.9,-0.05) .. controls (-1,0.8) and (1,0.8) .. (1.9,-0.05);
\draw[blue](0,0) ellipse  (3cm and 1.8cm);
\draw[red] (0,-2.975) arc (-90:90:0.65cm and 1.2cm);
\draw[red,dashed] (0,-0.575) arc (90:270:0.65cm and 1.2cm);
\scope[xshift=8.5cm,yshift=-2.2cm,scale=1.1]
\draw[->](-0.5,0) -- (7,0) node[above]{${\rm Re}(z)$};
\draw[->](0,-0.5) -- (0,4.5) node[left]{${\rm Im}(z)$};
\draw(0,0)node{$\bullet$};
\draw(-0.6,0.6)node{$0$};
\draw[blue](0,0) -- (1.5,3.5);
\draw (1.5,3.5)node{$\bullet$} ;
\draw(0.9,4.1)node{$\tau$};
\draw[red](0,0) -- (5,0);
\draw (5,0)node{$\bullet$};
\draw (4.4,0.6)node{$1$};
\draw[red](1.5,3.5) -- (6.5,3.5);
\draw[blue](5,0) -- (6.5,3.5);
\draw(6.5,3.5)node{$\bullet$};
\draw(5.9,4.1)node{$\tau{+}1$};
\endscope
\endtikzpicture
\caption{Parameterization of a torus as a lattice $\mathbb C/(\mathbb Z{+}\tau \mathbb Z)$ with 
discrete identifications $z \cong z{+}1 \cong z{+}\tau$ of the punctures 
and modular parameter $\tau$ in the upper half plane.}
\label{figureone}
\end{center}
\end{figure}

\vspace{-0.5cm}

\noindent
{\it \bf Open-string correlators}. Color-stripped one-loop amplitudes of $n$ open-string states 
are given by the moduli-space integral
\beq
A^{{\rm 1-loop}}_{\rm open}(\lambda) =
\int d^D\ell  \int_{D(\lambda)} d\tau \,
\prod_{j=2}^n dz_j \, |{\cal I}_n| \, {\cal K}_n \ .
\label{open}
\eeq
Following the {\it chiral-splitting} techniques of \cite{DHoker:1988pdl},
the integrations of (\ref{open}) involve $D$-dimensional loop momenta $\ell$.
The integration domain $D(\lambda)$ for the moduli $z_j,\tau$ depends on the
topology of the genus-one worldsheet---the cylinder and the M\"obius 
strip---represented by $\lambda$. Both of these topologies can
be derived from a torus via suitable involutions \cite{polchinski1}, and
its usual parametrization depicted in fig.\ \ref{figureone} requires the
quantities $|{\cal I}_n| \, {\cal K}_n$ to be doubly periodic functions 
as $z\rightarrow z{+}1$ and $z\rightarrow z{+}\tau$---at least after integration
over $\ell$.

{\it Koba--Nielsen factor and the correlators}.
A universal contribution to genus-one integrands in (\ref{open}) is furnished by the 
Koba--Nielsen factor $|{\cal I}_n|$ with 
\begin{equation}
{\cal I}_n \equiv \exp\Big( \sum^n_{i<j} s_{ij} \ln \theta(z_{ij},\tau) +  \sum_{j=1}^n  z_j(\ell\cdot k_j) + {\tau \ell^2 \over 4\pi i}  \Big)
\label{Koba}
\end{equation}
and $z_{ij} \equiv z_i{-}z_j$. Here, $s_{ij} \equiv k_i\cdot k_j$ are the Mandelstam invariants
in units $2\alpha'=1$ built from lightlike external momenta $k_j$, 
and $\theta$ is the odd Jacobi theta function
\begin{equation}
\theta(z,\tau) \equiv  \sin( \pi z)  \! \prod_{n=1}^{\infty}  \bigl( 1  {-}  e^{2\pi i (n \tau+z)} \bigr) \bigl( 1 {-} e^{2\pi i (n \tau-z)}\bigr) \, .
\label{oddth}
\end{equation}
Finally, the correlators ${\cal K}_n$ in (\ref{open}) are the main subject of this Letter's investigations:
They comprise kinematic factors for the external states written in pure-spinor superspace 
as well as meromorphic functions of the moduli to be introduced as generalized elliptic integrands. 
We will provide evidence via explicit examples at $n\leq 6$ points that the kinematic factors and 
generalized elliptic integrands satisfy identical relations and that their composition can be viewed as a double copy.

By virtue of chiral splitting, the moduli-space integrands of closed-string one-loop amplitudes 
follow as the holomorphic square ${\cal K}_n \rightarrow |{\cal K}_n |^2$ 
along with $|{\cal I}_n| \rightarrow |{\cal I}_n|^2$ \cite{DHoker:1988pdl}. Hence,
the double-copy structure to be described for ${\cal K}_n$ immediately propagates to the closed string.

{\it Kinematic factors from pure spinors}.
In the pure-spinor formulation of the superstring \cite{Berkovits:2000fe},
the gauge invariance and supersymmetry of the amplitudes
are unified to an invariance under the BRST operator $Q$.
A classification of BRST-invariant kinematic factors of various tensor
ranks that can arise from the one-loop amplitude prescription has
been given in \cite{Mafra:2014gsa}. The simplest scalar BRST
invariants can be expressed in terms of gauge-theory trees
\cite{Mafra:2014oia}, e.g.
\begin{align}
C_{1|2,3,4} &= s_{12}s_{23} A^{\rm tree}_{\rm YM}(1,2,3,4)\label{gaugetrees} \\
C_{1|23,4,5} &= s_{45} \big[ s_{34} A^{\rm tree}_{\rm YM}(1,2,3,4,5) - (2\leftrightarrow 3) \big]  \ , \notag
\end{align}
and further examples of various tensor ranks are available for download 
in \cite{website}. For instance, the bosonic components of a vector invariant $C^m_{1|2,3,4,5}$
(with Lorentz indices $m,n=0,1,\ldots,D{-}1$ in $D$ spacetime dimensions)
involve tensor structures such as $e_1^m t_8(2,3,4,5)$ and $\frac{k_2^m}{s_{12}} t_8(12,3,4,5)$.
The $t_8$ tensor with multiparticle insertions is defined in \cite{He:2016mzd}, and
$e_i$ denotes the polarization vector of the {\it i}th gluon. Here and in the 
following, groups of external-state labels in a
subscript that are separated by a comma (rather than a vertical
bar) can be freely interchanged, e.g.\ $C_{1|23,4,5} =C_{1|23,5,4}=C_{1|4,23,5}$.

In addition to BRST-invariant kinematic factors, the six-point
correlator \cite{Mafra:2016nwr} gives rise to {\it
pseudoinvariants} with nonvanishing BRST variations
\begin{align}
QC^{mn}_{1|2,3,4,5,6} &= - \eta^{mn} V_1 Y_{2,3,4,5,6} \notag \\
Q P_{1|2|3,4,5,6}&= - V_1 Y_{2,3,4,5,6} \,, \label{pseudo}
\end{align}
where $V_1$ denotes an unintegrated vertex operator
and $Y_{2,3,4,5,6}$ is related to the anomaly kinematic 
factor $\sim \varepsilon_{10}F^5$ 
of the gluon field strength \cite{Berkovits:2006bk}.
The BRST variation of the correlator localizes on the boundary 
of moduli space, and the cancellation of the hexagon anomaly 
\cite{Green:1984qs} thus follows as usual in the integrated 
amplitude (\ref{open}).

The construction of (pseudo)invariants from Berends--Giele currents \cite{Mafra:2014gsa}
gives rise to the shuffle symmetries within the individual groups of
labels, e.g.
\beq
C^{\ldots}_{1|23,\ldots} = -C^{\ldots}_{1|32,\ldots}  \ , \ \ \ \  
C^{\ldots}_{1|234,\ldots} +{\rm cyc}(2,3,4) = 0 \, .
\label{shuffl}
\eeq

{\it Double-copy representations}. In order to exemplify
the main result of this work, the correlators for open-string
amplitudes (\ref{open}) up to multiplicity six can be written as~\cite{wip}

\vspace{-0.4cm}

\begin{footnotesize}
\begin{align}
{\cal K}_4 &= C_{1|2,3,4}E_{1|2,3,4}\notag\\
{\cal K}_5 &= C^m_{1|2,3,4,5} E^m_{1|2,3,4,5} +  \big[ s_{23}C_{1|23,4,5} E_{1|23,4,5} {+} (2,3|2,3,4,5) \big]
\notag\\ 
{\cal K}_6&= \frac{1}{2} C^{mn}_{1|2,\ldots,6} E^{mn}_{1|2,\ldots,6}
 {-}  \big[ P_{1|2|3,4,5,6} E_{1|2|3,4,5,6}
{+} (2\leftrightarrow 3,\ldots,6) \big] \notag \\
& \! \! \! +  \big[ s_{23} C^m_{1|23,4,5,6} E^m_{1|23,4,5,6}
+ (2,3|2,3,4,5,6) \big] \label{KK4}\\
& \! \! \!   + \Big( \big[ s_{23} s_{45} C_{1|23,45,6} E_{1|23,45,6}
{+} {\rm cyc}(3,4,5) \big] + (6\leftrightarrow 5,4,3,2) \Big)  \notag
 \\
& \! \! \!  + \Big( \big[ s_{23}s_{34} C_{1|234,5,6}  E_{1|234,5,6}
{+} {\rm cyc}(2,3,4) \big]
{+} (2,3,4|2,\ldots,6) \Big) \, . \notag
\end{align}
\end{footnotesize}
Throughout this work, $(i_1,\ldots,i_p | i_1,\ldots,i_q)$ denotes a sum over the
${q \choose p}$ choices of $p$ indices $i_1,\ldots,i_p$ 
out of $ i_1,\ldots,i_q$. The entire dependence of the correlators (\ref{KK4}) on 
the external polarizations is captured by the above (pseudo)invariants, and they 
are accompanied by generalized elliptic integrands 
$E_{1|\ldots}^{\ldots}$ to be spelled out in the next section. In particular, the two kinds
of ingredients in (\ref{KK4}) will be shown to enter on completely 
symmetric footing and to be freely interchangeable. This symmetry is at the heart of 
the double-copy structure of one-loop open-string amplitudes.


\bigskip
{\it \bf Generalized elliptic integrands}.
At tree level, the double-copy structure of the open superstring
arises from a relation between kinematic factors and worldsheet
functions defined on a disk \cite{Broedel:2013tta}. The same Kleiss--Kuijf and
BCJ relations among gauge-theory amplitudes $A^{\rm tree}_{\rm
YM}(1,2,\ldots,n)$ \cite{Kleiss:1988ne, Bern:2008qjA} are satisfied
by the disk integrals of the so-called Parke--Taylor factors
$(z_{12} z_{23} \ldots z_{n-1,n} z_{n,1})^{-1}$, where $z_j$
represent the locations of the punctures on the disk boundary.

In this section, we will introduce the notion of {\it generalized elliptic integrands} (GEIs)
to specify the $E_{1|\ldots}^{\ldots}$ in the correlators (\ref{KK4}).
They refer to functions on genus-one Riemann surfaces that play a similar role in one-loop
open-string amplitudes as the Parke--Taylor factors at tree level.

{\it Key definition}.
By the quasiperiodicity $\theta(z{+}\tau,\tau)=- e^{-i\pi \tau -2\pi i z} \theta(z,\tau)$
of the theta function (\ref{oddth}), the Koba--Nielsen factor 
(\ref{Koba}) by itself is not a doubly periodic
function of the punctures. However, its monodromies as $z_j \rightarrow z_j{+}\tau$
can be compensated by a shift in the loop momentum 
$\ell \rightarrow \ell {-}2\pi i k_j$:
\begin{equation}
{\cal I}_n \big|_{z_j \rightarrow z_j {+} \tau}^{\ell \rightarrow \ell {-}2\pi i k_j}  = {\cal I}_n 
\, .
\label{defGEF}
\end{equation}
We refer to meromorphic functions of $z_j,\ell,\tau$ invariant under $(z_j,\ell)
{\rightarrow} (z_j {+} \tau, \ell {-}2\pi i k_j)$ and $(z_j,\ell) {\rightarrow} (z_j {+} 1, \ell)$
as GEIs. After integrating the loop momentum in 
(\ref{open}), GEIs give rise to doubly periodic but generically nonmeromorphic functions of $z_j$ and $\tau$.
Since ${\cal I}_n$ transforms by a complex phase under $z_j \rightarrow z_j {+} 1$, the
quantity $|{\cal I}_n|$ in (\ref{open}) is a GEI.

{\it Scalar GEIs}. A variety of GEIs can be generated from the Kronecker--Eisenstein series \cite{Kroneck},
\begin{equation}
F(z,\alpha,\tau) \equiv { \theta'(0,\tau) \theta(z{+}\alpha,\tau) \over  \theta(\alpha,\tau)  \theta(z,\tau)} \equiv \sum_{n=0}^{\infty} \alpha^{n-1} g^{(n)}(z,\tau) 
\label{Kron}
\end{equation}
whose expansion in $\alpha$ defines meromorphic functions such as
$g^{(0)}(z,\tau)=1$ and $g^{(1)}(z,\tau) = \partial_z \ln
\theta(z,\tau)$ as well as
\beq
2g^{(2)}(z,\tau) {=} 
  (\partial_z \ln \theta(z,\tau))^2 {+} 
	       \partial_z^2  \ln \theta(z,\tau) {-} 
	      {\theta'''(0,\tau)  \over  3 \theta'(0,\tau)}  \ .
\eeq
The importance of the Kronecker--Eisenstein series
to the description of one-loop open-string integrands has been
recently emphasized in \cite{eMZV}, where it was shown to reproduce
the spin-sum identities of \cite{Tsuchiya:1988va}.

The quasiperiodicity $F(z{+}\tau,\alpha,\tau)=e^{-2\pi i\alpha}
F(z,\alpha,\tau)$ implies that the functions
$g^{(n)}(z,\tau)$ are not elliptic,
\beq
\label{monodromy}
g^{(n)}(z+\tau,\tau) = \sum_{k=0}^n \frac{(-2\pi i)^k}{k!}g^{(n-k)}(z,\tau)\,,
\eeq
for example,
$g^{(1)}(z{+}\tau,\tau)=g^{(1)}(z,\tau)-2\pi i$. However, these
monodromies cancel in cyclic products
\begin{align}
&F(z_{12},\alpha,\tau) F(z_{23},\alpha,\tau) \ldots F(z_{k-1,k},\alpha,\tau) F(z_{k,1},\alpha,\tau)
\notag \\
&\equiv \alpha^{-k}\sum_{w=0}^{\infty} \alpha^w V_w(1,2,\ldots,k) \label{prodKron}
\end{align}
which define elliptic functions $V_w$ in $k$ variables with $w$
simultaneous poles as $z_j \rightarrow z_{j+1}$ such as $V_0(1,2,\ldots,k){=}1$ and
$V_1(1,2,\ldots,k)=\sum_{j=1}^k g^{(1)}_{j,j+1}$ as well as
\begin{align}
&V_2(1,2,\ldots,k)=\sum_{j=1}^k g^{(2)}_{j,j+1} {+}\sum_{i<j}^k g^{(1)}_{i,i+1} g^{(1)}_{j,j+1}
\end{align}
with $g^{(n)}_{ij} \equiv g^{(n)}(z_i{-}z_j,\tau)$ and $z_{k+1}\equiv z_1$.
Therefore, the following functions are elliptic:
\begin{align}
E_{1|23,4,5} &\equiv V_1(1,2,3) \ , \ \ \ \ \ \ 
E_{1|234,5,6} \equiv V_2(1,2,3,4) \notag \\
E_{1|23,45,6} &\equiv V_1(1,2,3) V_1(1,4,5)   \ .
\label{boildown}
\end{align}
In addition to the above scalar elliptic functions, we also introduce
the formal definition $E_{1|2,3,4}\equiv1$ and
\begin{equation}
\label{Pell}
E_{1|2|3,4,5,6} \equiv \partial_{z_1} g^{(1)}_{12} + s_{12} (g^{(1)}_{12})^2 - 2 s_{12} g^{(2)}_{12} 
\end{equation}
which exhaust the scalar GEIs in the correlators (\ref{KK4}).

{\it Tensorial GEIs}.
Open-string integrands at $(n\geq 5)$
points also involve loop momenta from the zero modes of certain
worldsheet fields. Appearances of $\ell$ will be combined with the
coefficients $g^{(n)}$ of the Kronecker--Eisenstein series
(\ref{Kron}) to form GEIs such as
\begin{align}
E^m_{1|2,3,4,5} &\equiv \ell^m + k_2^m g^{(1)}_{12}
+ k_3^m g^{(1)}_{13}
+ k_4^m g^{(1)}_{14}
+ k_5^m g^{(1)}_{15}\notag\\
E^m_{1|23,4,5,6} &\equiv\big( \ell^m {+} k_4^m g^{(1)}_{14}{+} k_5^m g^{(1)}_{15}{+} k_6^m g^{(1)}_{16}\big)V_1(1,2,3) \notag \\
& \! \! \! \! \!  \! \! \! \! \! \! \! \! \! + \big[k_2^m( \gg1(1,3)\gg1(2,3) 
          {+} \gg2(1,2)
          {-} \gg2(1,3)
          {-} \gg2(2,3)) - (2\leftrightarrow 3)\big] \label{E6vec}\\
E^{mn}_{1|2,3,4,5,6} &\equiv \ell^m\ell^n + \big[k_2^{(m}k_{3}^{n)}\gg1(1,2)\gg1(1,3)+
(2,3|2,\ldots,6)\big]
\notag\\
& \! \! \! \! \!  \! \! \! \! \! \! \! \! \!   +\big[\ell^{(m}k_2^{n)}\gg1(1,2) +2 k_2^m k_2^n g^{(2)}_{12}
 + (2\leftrightarrow3,4,5,6)\big] \ .\notag
\end{align}
Vector indices
are {symmetrized} according to 
$\ell^{(m}k_2^{n)}=\ell^{m}k_2^{n}{+}\ell^{n}k_2^{m}$,
and the notation for the permutations is explained below
(\ref{KK4}).

One can explicitly check that the above $E^{\ldots}_{1|\ldots}$
constitute GEIs after using \eqref{monodromy} and momentum
conservation. These GEIs suffice to describe open-string correlators 
(\ref{KK4}) up to six points, and higher multiplicities or tensor ranks 
will be addressed in \cite{wip}.

{\it Shuffle symmetries from Fay identities}. Similar to the kinematic factors, 
GEIs obey shuffle symmetry within the individual groups of labels; e.g.,
\beq
E^{\ldots}_{1|23,\ldots} = -E^{\ldots}_{1|32,\ldots}  \ , \ \ \ \  
E^{\ldots}_{1|234,\ldots} +{\rm cyc}(2,3,4) = 0 \, .
\label{GEIshuffl}
\eeq
These shuffle symmetries can be traced back to the symmetry $g^{(n)}_{ij}=(-1)^n g^{(n)}_{ji}$
and the components of the Fay relations \cite{Kroneck}
\beq
F(z_1,\alpha_1)F(z_2,\alpha_2) =
F(z_1,\alpha_1{+}\alpha_2) F(z_2{-}z_1,\alpha_2) {+}(1\leftrightarrow 2)
\label{Fay}
\eeq
such as \cite{eMZV} $g^{(1)}_{12} g^{(1)}_{23} + g^{(2)}_{13}
+ {\rm cyc}(1,2,3) = 0$.


\bigskip
\noindent
{\it \bf The double-copy structure}.
In this section we will show surprising relations between the BRST-invariant 
kinematic factors and GEIs that underpin the double-copy
structure of the open superstring at one loop.
When trading the GEIs in the correlators \eqref{KK4}
for another copy of kinematic factors, gravitational
matrix elements of $R^4$ will be seen to emerge.


{\it BRST-invariant kinematic factors versus GEIs}.
Given the GEIs defined above, one can show that
\beq
k_2^m E^m_{1|2,3,4,5} + \big[ s_{23} E_{1|23,4,5} {+} (3\leftrightarrow 4,5)
\big] = 0  
\label{ellid}
\eeq
up to a total worldsheet derivative $\frac{\partial  \ln {\cal I}_5}{\partial z_2}$
that vanishes under the integrals of (\ref{open}).
Rather surprisingly, in 2014 the following kinematic identity
of identical structure was proven in the cohomology of the BRST operator
\cite{Mafra:2014gsa},
\beq
\label{k2Cm}
k_2^m C^m_{1|2,3,4,5} + \big[ s_{23} C_{1|23,4,5} {+}
(3\leftrightarrow 4,5)\big] = 0\, ,
\eeq
as can be explicitly verified with the data provided on the website
\cite{website}. Note that (\ref{k2Cm}) enters the field-theory amplitudes
of \cite{Mafra:2014gja} as a kinematic Jacobi identity \cite{Bern:2008qjB}.

The striking resemblance between the identities
\eqref{ellid} on a genus-one Riemann surface and \eqref{k2Cm} in
the cohomology of the kinematic BRST operator
motivates us to search for further instances. 
Indeed, the symmetry properties~\cite{Mafra:2014gsa}
\begin{align}
C_{2|34,1,5} &= C_{1|34,2,5} + C_{1|23,4,5} - C_{1|24,3,5} \notag \\
C_{2|13,4,5} &=  - C_{1|23,4,5} \label{canonicalize} \\
C^m_{2|1,3,4,5} &= C^m_{1|2,3,4,5} + \big[ k_3^m C_{1|23,4,5} + (3\leftrightarrow 4,5) \big]  \notag
\end{align}
hold for their dual GEIs in identical form
\begin{align}
E_{2|34,1,5} &= E_{1|34,2,5} + E_{1|23,4,5} - E_{1|24,3,5} \notag \\
E_{2|13,4,5} &=  - E_{1|23,4,5} \label{canonicalize2} \\
E^m_{2|1,3,4,5} &= E^m_{1|2,3,4,5} + \big[ k_3^m E_{1|23,4,5} + (3\leftrightarrow 4,5) \big] \ , \notag
\end{align}
as can be verified from their explicit expressions above.
Similarly, the kinematic identities at six points \cite{Mafra:2014gsa}
\begin{align}
k_{23}^m C^{m}_{1|23,4,5,6} &= P_{1|2|3,4,5,6}   - P_{1|3|2,4,5,6}    \label{kinatsixa}  \\
 + \big[ s_{24} &C_{1|324,5,6} - s_{34} C_{1|234,5,6} + (4\leftrightarrow 5,6) \big]
\notag \\
k_1^m C^{mn}_{1|2,3,4,5,6} &=- \big[ k_2^n P_{1|2|3,4,5,6}  + (2\leftrightarrow 3,4,5,6) \big]  \label{kinatsix}   \\
\eta_{mn} C^{mn}_{1|2,3,4,5,6} &= 2  \big[ P_{1|2|3,4,5,6}  + (2\leftrightarrow 3,4,5,6) \big]   \label{kinatsixb} 
\end{align}
all have a direct counterpart in terms of GEIs up to boundary terms in moduli
space. Under $C^{\ldots}_{1|\ldots} \rightarrow E^{\ldots}_{1|\ldots}$, (\ref{kinatsixa}) 
and (\ref{kinatsix}) translate into total derivatives with respect to the punctures $z_j$ that can
be immediately discarded. The GEI-analogue of (\ref{kinatsixb}) additionally involves 
a $\tau$ derivative
\beq
\eta_{mn} E^{mn}_{1|2,\ldots,6} = 2  \big[ E_{1|2|3,4,5,6}  {+} (2\leftrightarrow 3,\ldots,6) \big]  +4\pi i \frac{\partial  \ln {\cal I}_6}{\partial \tau} \, ,
\label{taude}
\eeq
resulting in the expected BRST anomaly
$Q({\cal K}_6 {\cal I}_6) \sim \frac{\partial  {\cal I}_6}{\partial \tau} $.

The above identities among GEIs can be derived from the antisymmetry 
$g^{(1)}_{ij} = - g^{(1)}_{ji}$, momentum conservation, and the Fay
identity (\ref{Fay}).
Together with the shuffle symmetries (\ref{shuffl}), these relations signal a
fascinating {\it duality between BRST invariants and GEIs} which will be
shown on a case-by-case basis to persist at higher points \cite{wip}. 

The duality even extends to anomalies: The 
BRST variations (\ref{pseudo}) can be mapped to a modular anomaly in the 
$\ell$ integral over $E^{mn}_{1|2,\ldots,6}$ and $E_{1|2|3,4,5,6} $ \cite{wip}
which cancels by the kinematic identity (\ref{kinatsixb}) dual to (\ref{taude}).


{\it Comparison with $R^4$}. In the low-energy limit, one-loop amplitudes 
of the closed string are known to yield matrix elements of higher-curvature
operators $R^4$ \cite{Green:1982sw}. Up to and including six points, they have been
expressed in terms of the above BRST (pseudo)invariants \cite{Mafra:2016nwr},

\vspace{-0.3cm}
\begin{footnotesize}
\begin{align}
{\cal
M}_4^{R^4}&= C_{1|2,3,4}\tilde C_{1|2,3,4}
\notag \\
{\cal M}_5^{R^4}&= C^m_{1|2,3,4,5} \tilde C^m_{1|2,3,4,5}
+  \big[ s_{23}C_{1|23,4,5} \tilde C_{1|23,4,5} {+} (2,3|2,3,4,5) \big]  \notag
\\
{\cal M}_6^{R^4}&= \frac{1}{2} C^{mn}_{1|2,\ldots,6} \tilde C^{mn}_{1|2,\ldots,6}  {-}  \big[ P_{1|2|3,4,5,6} \tilde P_{1|2|3,4,5,6}  {+}(2 {\leftrightarrow} 3,\ldots,6) \big] \notag \\
& \! \! \! +  \big[ s_{23} C^m_{1|23,4,5,6} \tilde C^m_{1|23,4,5,6} + (2,3|2,3,4,5,6) \big] \label{R46} \\
& \! \! \!   + \Big( \big[ s_{23} s_{45} C_{1|23,45,6} \tilde C_{1|23,45,6} {+} {\rm cyc}(3,4,5) \big] + (6\leftrightarrow 5,4,3,2) \Big)  \notag
 \\
& \! \! \!  + \Big( \big[ s_{23}s_{34} C_{1|234,5,6}  \tilde C_{1|234,5,6}{+} {\rm cyc}(2,3,4) \big]    {+} (2,3,4|2,\ldots,6) \Big) \, . \notag  
\end{align}
\end{footnotesize}
\noindent The tilde refers to a second copy of the
superspace kinematic factors, where the gravitational polarizations
can be reconstructed from the tensor product of the gauge-theory
polarizations. The double-copy structure of the above ${\cal M}_n^{R^4}$
is shared by the open-string correlators \eqref{KK4} which are converted
to (\ref{R46}) by trading the GEIs for another copy of their kinematical correspondents: 
$E\leftrightarrow \{\tilde C,\tilde P\}$. This motivates us to conjecture that
\beq
{\cal K}_n = {\cal M}_n^{R^4} \, \big|_{\tilde C,\tilde P \rightarrow E} 
\label{allmult}
\eeq
for arbitrary multiplicities $n$, where all the vector indices and external-particle labels in the
subscripts are understood to be inert under the replacements.
At multiplicity $n=7$, \eqref{allmult} leads to a new 
supersymmetric expression for ${\cal K}_{7}$,
i.e.\ the results of this Letter allow us to
probe uncharted terrain of multiparticle string amplitudes 
(see \cite{wip} for details and consistency checks).


\bigskip
{\it \bf Conclusions and outlook}.
In this Letter, we have presented evidence for a duality
between GEIs and BRST-invariant kinematic factors: identities
among GEIs that vanish up to boundary terms in moduli space are mapped to
identities among kinematic factors that vanish up to BRST-exact
terms. This duality has been exploited to reveal a double-copy structure in
the one-loop amplitudes of the open superstring. Trading
GEIs in {\it open-string} correlators by another copy of
BRST-invariant kinematic factors leads to {\it gravitational}
matrix elements of supersymmetrized $R^4$ operators.

The duality between elliptic functions and BRST invariants presented
here turns out to be even richer. Alternative double-copy
representations of the above open-string correlators will be given 
in \cite{wip} which manifest their locality instead of gauge
invariance. These representations will illustrate further aspects of
the duality between kinematic factors and worldsheet functions, in
closer contact with conformal-field-theory techniques.

It is a fascinating possibility that the duality between
kinematic invariants and (generalized) elliptic functions is a
generic feature of string-theory correlators.
At genus $g=2,3$, the low-energy limits of closed-string
amplitudes have been recently computed with the pure-spinor
formalism, resulting in matrix elements of $D^{2g}R^4$
\cite{higherloop}. It is conceivable that their double-copy
structure applies to open-string correlators at the respective loop
order.

The new double-copy structures unraveled in this Letter should lead to great 
simplification of higher-order calculations in string theory, deducing the structure
of the integrands from effective-field-theory quantities. Moreover, the study of GEIs is 
expected to trigger conceptual advances in the mathematics of string theory related 
to the interplay of higher-genus geometry and algebra. Finally, the $\alpha'\rightarrow 0$
limit of our string-theory results yields new representations of field-theory amplitudes 
and will shed further light on the BCJ double copy at loop level~\cite{anotherwip}.

\medskip \noindent\textit{Acknowledgements:} We are grateful to
Freddy Cachazo and in particular Henrik Johansson for valuable comments 
on a draft. We are indebted to the IAS Princeton
and to Nima Arkani-Hamed for kind hospitality during an inspiring
visit which initiated this project.
This research was supported by the Munich Institute for Astro- and Particle
Physics (MIAPP) of the DFG cluster of excellence ``Origin and
Structure of the Universe''. CRM is supported by a University 
Research Fellowship from the Royal Society.
The research of OS was supported in part by 
Perimeter Institute for Theoretical
Physics. Research at Perimeter Institute is supported by the Government of
Canada through the Department of Innovation, Science and Economic Development
Canada and by the Province of Ontario through the Ministry of Research,
Innovation and Science.

\end{document}